# The electronic structure of cuprates from high energy spectroscopy

Mark S. Golden*, Christian Dürr, Andreas Koitzsch, Sibylle Legner, Zhiwei Hu, Sergey Borisenko, Martin Knupfer and Jörg Fink

*Institute for Solid State Research, IFW Dresden*
*P.O. Box 270016, D-01171 Dresden, Germany*

*Corresponding author. Email address: m.golden@ifw-dresden.de

Abstract

We report studies of the electronic structure and elementary excitations of doped and undoped cuprate chains, ladders and planes. Using high energy spectroscopies such as x-ray absorption, core level photoemission and angle resolved photoemission spectroscopy, important information regarding the charge distribution and hole dynamics can be obtained. The comparison of the experimental data with suitable theoretical models sets constraints on the parameters entering into the model calculations, and offers insight into the important physical quantities governing the electronic structure of these materials. Recurring themes include the importance of the dimensionality of the Cu-O network (1D→2D) and the crucial role played by the spin-background in determining the dynamics of low-lying excitations in these strongly correlated systems.

## 1. Introduction

The discovery of high temperature superconductivity in the late 80's has made the cuprates the most intensively studied class of substances in solid state research. Not only the high transition temperatures make the cuprates fascinating, but also their far from conventional normal state properties. Two main ingredients go into making the cuprate dish so interesting and - from a physics point of view – so nourishing: electronic correlations and low-dimensionality. These are certainly the cause of much of the cuprates' extraordinary physical properties in the normal state and furthermore many believe that they play a defining role in the mechanism of high temperature superconductivity itself.

The source of the dominating role played by correlation effects in the cuprates is simple: in the divalent ground state (the starting point for all cuprate materials considered here) each Cu ion possess nine 3d electrons – i.e. the 3d shell contains a single hole. The compact nature of the 3d orbitals means that two electrons, or holes, repel each other if they are forced to reside at the same site. In the cuprates, this Coulomb interaction energy, $U$, is of the order of 10 eV. As the energy reduction resulting from delocalisation of the electrons (determined by the band width) is less than $U$, a suppression of the charge fluctuations required for electrical conductivity ($Cu3d^9 + Cu3d^9 \rightarrow Cu3d^8 + Cu3d^{10}$) is the consequence. Thus, even though in a simple one electron picture an *undoped* cuprate already has a half-filled (antibonding $dp_\sigma$) band – from overlap of the $Cu3d_{x^2-y^2}$ and $O2p_{x,y}$ orbitals – in reality it is an insulator with an energy gap of some 2 eV. The fact that the energy gap is much smaller than U results form the presence of the O2p valence band states between the correlation-split lower and upper Hubbard bands, making the cuprates charge transfer insulators[1]. Seen against the backdrop of the continued search for a broadly-accepted mechanism describing the high temperature superconductors (HTSC's), understanding the influence of electronic correlation in the cuprates remains one of the oustanding challenges in solid state physics.

Correlation effects are most sensitively felt when the system undergoes excitation, as for example is the case in electronic transport. High energy spectroscopies possess the advantage that they deal exclusively in excitations of the electron system, and are therefore ideally suited to the investigation of correlation effects in solids. Here we will discuss results stemming from two classes of experiments – x-ray absorption spectroscopy and photoemission spectroscopy. The former can, under certain conditions be envisaged as an electron addition spectroscopy (N $\rightarrow$ N+1), whereas the latter involves the study of the system after removal of an electron (N $\rightarrow$ N-1). The combination of both methods gives access to both the low and high energy scales in the cuprates, either side of the chemical potential.

There exist a wide variety of different one-dimensional (1D) and two-dimensional (2D) copper-oxygen networks built up of square-planar $CuO_4$ plaquettes, the best known representatives of which are the essentially 2D high-temperature superconductors (HTSC). Parallel to the study of the highly complex HTSC themselves, much can be learned regarding the physics of the cuprates by investigating the elementary excitations of model (often undoped) cuprate networks using high energy spectroscopy. In the 2D systems, information of direct relevance to the low-doping regime of the HTSC can be gathered, while the 1D networks offer direct experimental access to low dimensional, correlated electron systems, which offer a rich physics worthy of detailed investigation in its own right.

In this contribution, we give a brief treatment of the kind of insight that can be gained into the electronic structure of cuprates (from 1D through 1½D to 2D) from electron spectroscopic and related techniques. We cannot attempt to provide a rigorous, in-depth review of this topic, but do hope to be able to convey the complementary nature of the information available when a variety of high energy spectroscopies such as the photoemission of core and valence levels and x-ray absorption are systematically applied to a series of structurally related yet subtly differing cuprate networks. In order to cover the breadth of compounds and different methods in the space provided some of the results presented, and in particular some of the theory, will have to be accepted at face value.

This special issue of the journal contains a number of other articles which are complementary to this contribution, including notably the paper by J. Fink *et al.* covering the study of correlation effects in solids (including cuprates) using electron energy-loss spectroscopy as well as the papers by Z.-X. Shen *et al.* and J.-C. Campuzano *et al.* regarding the electronic structure of the HTSC's from ARPES.

This paper is structured as follows: in the experimental section, after introducing the Cu-O networks under investigation, a brief mention will be given of the spectroscopic techniques and conditions involved. To ease discussion, the results are then presented and discussed grouped into sections according to the spectroscopy concerned (x-ray absorption: XAS; core level photoemission: XPS; angle resolved photoemission: ARPES). The paper closes with brief concluding remarks and an outlook.

## 2. Experimental

*Model cuprates and Cu-O network structures.*

The almost unparalleled architectural flexibility offered by the square-planar $CuO_4$ plaquette was already exploited in the synthesis of novel cuprate networks in inorganic chemistry in the seventies[2], without it being realised what a potential such systems offer for the systematic variation of the important electronic and magnetic interactions in such model late transition metal oxides. Fig. 1 gives a schematic illustration of the Cu-O networks dealt with in the course of this paper. Of particular importance in each structure is the nature of the electronic or magnetic pathway joining neighbouring Cu ions[3], which depends sensitively upon the manner in which the $CuO_4$ plaquettes are joined to one another. A 180° Cu-O-Cu pathway represents a strong magnetic link and results in a large positive (antiferromagnetic, AFM) exchange integral, J, whereas a 90° Cu-O-Cu pathway represents a weak coupling of one Cu ion to the next with the possibility of a negative (ferromagnetic, FM) nearest neighbour exchange integral. Thus a wide variety of cuprates can be conceived which contain Cu-O networks whose 'low dimensionality' stems from the Cu-O-Cu interaction pathway of the particular Cu-O system in question.

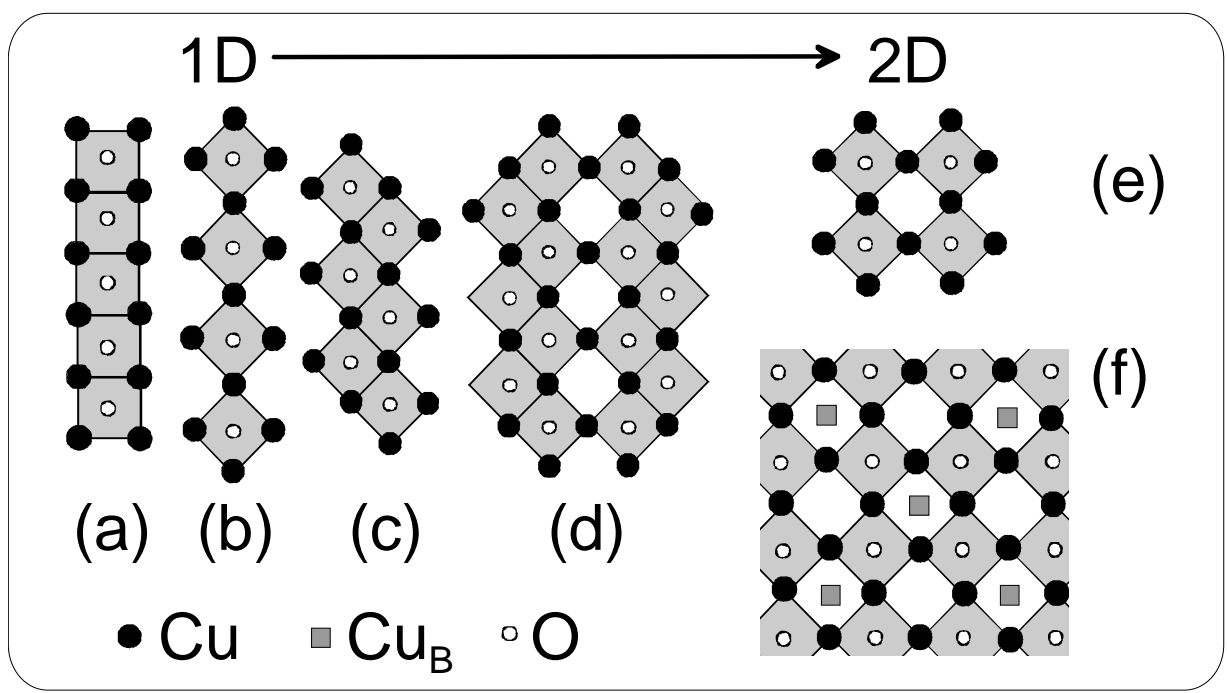

*Fig. 1.*
*Schematic representations of the Cu-O networks discussed in this paper. (a) edge-shared $CuO_2$ chain; (b) corner shared $CuO_3$ chain; (c) zig-zag chain; (d) two-leg ladder; (e) $CuO_2$ plane and (f) $Cu_3O_4$ plane. In each case the $CuO_4$-plaquettes are indicated by grey shading.*

The first Cu-O network depicted in Fig. 1a is that of a series of $CuO_4$ plaquettes coupled by their edges, giving a $CuO_2$ chain. This network can be found, for example, in $Li_2CuO_2$ [Cu(II)][4] and $NaCuO_2$ [Cu(III)], as well as in the inorganic spin-Peierls system $GeCuO_3$ [Cu(II)]. Following from the discussion above, an ideal 'edge-shared' chain, such as that depicted in Fig. 1a, has a 90° Cu-O-Cu interaction pathway, meaning that it is not a quasi-one-dimensional electronic system since the interplaquette coupling is so weak. Thus $Li_2CuO_2$ and its structural analogues are model systems for isolated $CuO_4$ plaquettes, i.e. these represent quasi zero-dimensional systems.

Fig. 1b depicts an alternative arrangement of $CuO_4$ plaquettes, now connected via their corners. $Sr_2CuO_3$ contains such 'corner-shared' $CuO_3$ chains[5], in which the 180° Cu-O-Cu interaction pathway means that this system is the best realisation of a one-dimensional quantum antiferromagnet found to date.

Fig. 1c shows the so-called zig-zag or double-chain system, which can be thought of as resulting from joining two corner-shared chains along their edges. $SrCuO_2$ is an example of a system which contains such double chains[6]. Due to the 180° Cu-O-Cu interaction pathway along each single chain, and the 90° pathway from the Cu of one chain to its neighbour in the second chain, such double chain systems behave in the first approximation electronically and magnetically speaking as a pair of separate single 1D chains, although the presence of the second chain does result in suppression of long range order due to frustration.

Fig. 1d illustrates the next step in the building block scheme, in which two double-chains are attached to one another. This results in a two-leg ladder, which is part of a large family of systems in which the strong 180° Cu-O-Cu interaction pathways form ladder-like structures, with 'legs' running along the chain axes and 'rungs' between the chain blocks[7]. The magnetic and electronic properties of these systems show a fascinating dependence on the number of legs in the ladder structure – with the ladder family being split into members containing either an even or odd number of legs[7].

Fig. 1e shows the most familiar of all the cuprate networks – the 2D $CuO_2$ plane – which forms the essential building block of all superconductors with critical temperatures exceeding 77K, the boiling point of liquid nitrogen. In its undoped form the $CuO_2$ plane is a model 2D quantum antiferromagnet.

Fig. 1f illustrates a 2D $Cu_3O_4$ plane as is contained in the model cuprate $Ba_2Cu_3O_4Cl_2$ [2], which is comprised of a $Cu_AO_2$ subsystem (as that shown in Fig. 1e), with additional $Cu_B$ sites (depicted in the figure as grey squares). The two subsystems together can be envisaged as a pair of interpenetrating 2D Heisenberg antiferromagnets, which possess quite different Néel temperatures: $T_N^A$=330 K, $T_N^B$=31 K [8]. The coupling between the A and B subsystems is weak and mainly due to quantum spin fluctuations, since it is based upon a 90° $Cu_A$-O-$Cu_B$ coupling and since the coupling of $Cu_A$ to $Cu_B$ is frustrated.

Please note that in the rest of the paper we label the crystallographic axes of all the cuprate systems dealt with in a unified manner: in each case the a,b plane contains the Cu-O network concerned, with the a axis representing the 1D direction of chain or ladder units; the c-axis is perpendicular to the $CuO_4$ plaquettes. This notation is not the same as that adopted in some of the original papers concerned with the structure of these materials, but simplifies the discussion and avoids confusion.

*High energy spectroscopies.*

In order to investigate the charge distribution, the charge carrier dynamics and elementary excitations in the cuprate networks sketched in Fig. 1, different high energy spectroscopies were employed.

X-ray absorption spectroscopy (XAS) was carried out using linearly polarised synchrotron radiation taken either from the SX-700-II monochromator operated by the Freie Universität Berlin at BESSY or from the U4B beamline at the NSLS (Brookhaven). The spectra were recorded at either the O-K or Cu-$L_3$ absorption edges using fluorescence detection or total electron yield, respectively, with energy resolutions in the range of 220-280 meV (O-K) and 650 meV (Cu-$L_3$). The spectra are normalised some 60 eV above the absorption threshold where the signal is polarisation independent. Core-level photoemission measurements were carried out using a commercial spectrometer equipped with a monochromatised Al-K$\alpha$ source, giving a total energy resolution of 400 meV. Angle-resolved photoemission (ARPES) measurements were carried out using commercial spectrometer systems at various beamlines at BESSY [U2-FSGM, 2m-Seya] (Berlin) and HASYLAB [Winkelemi, Honormi] (DESY, Hamburg) with an angular resolution of ±1° and an energy resolution of between 60 and 80 meV. The laboratory-based angle-scanned photoemission experiments were carried out using a commercial analyser (SCIENTA SES200), coupled to a monochromatised high performance VUV source (Gammadata VUV5000) with a total energy and angular resolution of 30 meV and ±0.3°. For the angle-scanned ARPES measurements the samples were mounted on a purpose-built manipulator-cryostat developed in our institute to provide a very precise rotation ($\Delta\theta$=0.1°) of the sample around three axes for temperatures down to 30 K. However, all measurements presented here were measured at or near to room temperature. For the XAS measurements, single crystalline samples were either cleaved in-situ or cut prior to measurement using the diamond knife of an ultramicrotome; polycrystalline samples were scraped in-situ using a diamond file. For ARPES, single crystals were cleaved in situ.

### 3. X-ray absorption spectroscopy of model 1D cuprate chains and cuprate ladders

*The edge-shared and corner-shared Cu-O chain systems: $Li_2CuO_2$ and $Sr_2CuO_3$*

Fig. 2 shows the polarisation dependent XAS spectra of $Li_2CuO_2$ and $Sr_2CuO_3$ recorded at the O-K edge. As discussed in Section 2, these two compounds contain the Cu-O chain building blocks that form the basis of many cuprate planes and ladders. The simplicity of the Cu-O networks in these systems enables an orbital-resolved stock-take of the intrinsic hole density in the O2p and Cu3d orbitals using polarisation dependent XAS.

We will mainly focus on the peak directly above the absorption onset, which is related to transitions into $O2p_{x,y}$ orbitals hybridised with $Cu3d_{x^2-y^2}$ states forming the upper Hubbard band or UHB[9]. As in all other cuprates, the O-K edges and Cu-$L_3$ spectra (not shown), indicate that the UHB is nearly completely built up from in-plaquette orbitals[9]; the spectral weight of unoccupied out-of-plane $O2p_z$ states in Fig. 2 (E||c) and of empty $Cu3d_{3z^2-r^2}$ states is small[10].

Since the in-plaquette hole states at the Cu sites have $3d_{x^2-y^2}$ character, no anisotropy of the Cu-$L_3$ absorption edges in the a,b plane (not shown) is observed, as expected. Turning now to the O-K spectra, for the edge-shared chain in $Li_2CuO_2$, the in-plaquette spectra (Fig. 2a) are practically identical in intensity, but, as the inset shows in more detail, do not occur at exactly the same energy, with the spectrum for E||b lying some 150 meV at lower energy. There are a number of possible reasons for this[11], which include: excitation into different parts of the Cu3d/O2p hybridised UHB depending upon whether the final state involves $O2p_x$ or $O2p_y$ orbitals; anisotropy in the coupling of phonons into the core level excitation process itself and correlation effects in the XAS final state. However, the energy shift is a detail which does not essentially change the isotropic nature of the in-plaquette O-K pre-edge structures in the XAS of a corner-shared chain – in agreement with the simple consideration that this structure only has one inequivalent O site.

For the O-K spectra of the corner-shared chain in $Sr_2CuO_3$ (Fig. 2b), the situation changes. Now two different low-lying peaks are observed in the a,b plane XAS[10], a fact which follows naturally from the two inequivalent O sites in the chain structure. The energetically lower-lying peak is due to transitions into unoccupied $O2p_y$ states of the two peripheral O(2) sites and the higher lying peak corresponds to unoccupied $O2p_x$ states of the central O(1) sites connecting the plaquettes[12]. The difference in energy between the two peaks may be explained by different Madelung potentials acting on the O1s levels and/or the UHB states of the two O sites.

The striking difference with respect to the edge-shared chain regards the intensities of the two pre-edge structures in $Sr_2CuO_3$. The O(1) site has two Cu neighbours compared to a single Cu neighbour for each of the O(2) sites. Correspondingly, as the O-K XAS intensity of the lowest lying structure is an expression of the hybridisation with the Cu orbitals forming the UHB, in the simplest picture one would expect twice the hole occupation for O(1) compared with O(2). Upon analysis of the peak areas, however, one finds a ratio for the hole occupation numbers, n[O(1)]/ n[O(2)] ~ 0.6 instead of 0.5. Thus, holes are pushed out from the central O(1) sites to the peripheral O(2) sites[12]. This can be explained by a non-zero intersite Coulomb interaction $V_{pd}$ (the central O(1) atom with its 2 Cu neighbours experiences this interaction more strongly), and/or by a difference of the charge-transfer energy $\Delta_{pp} = \Delta_{O(2)} - \Delta_{O(1)}$ for the two O sites, resulting from different Madelung potentials at the two O sites. A more quantitative analysis[12] gives $V_{pd}$ ~ 2-2.5 eV and $\Delta_{pp}$ ~ 0.5-1 eV.

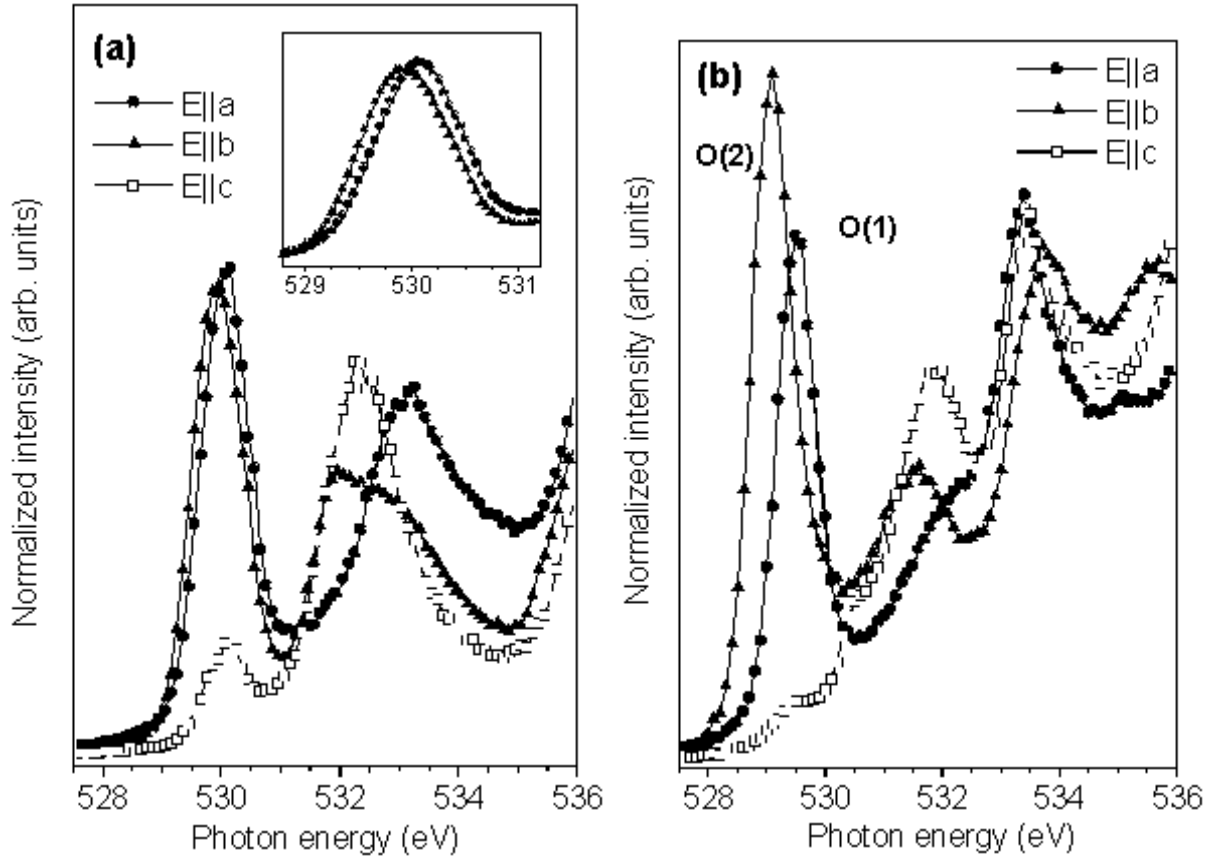

Fig. 2.
Polarisation dependent XAS of (a) $Li_2CuO_2$ [12] and (b) $Sr_2CuO_3$ [13] taken at the O-K pre-edge region. The inset in (a) shows the leading feature for E||a and E||b [11] on an expanded scale.

Similar measurements[13] have been performed on the zig-zag chain $SrCuO_2$ (see Fig. 1d). This case is a little more complicated as the central O sites have not only $O2p_x$ orbitals σ-hybridised with Cu sites along the chain direction, but also σ-hybridised $O2p_y$ orbitals perpendicular to the chain direction. However, the basic picture is the same: the hole density is pushed towards the periphery of the chains, meaning that upon taking a reasonable value for $\Delta_{pp}$, $V_{pd}$ values emerge[12] for $SrCuO_2$ which are analogous to those observed in $Sr_2CuO_3$.
These examples show that valuable information regarding the magnitude of important model parameters, such as those describing the off-site Coulomb repulsion energy, $V_{pd}$ can be obtained from such measurements. It appears that the reduction from 2D to 1D causes a significant increase in $V_{pd}$ from values <1 eV for $CuO_2$ plane systems[14] to 2-3 eV in the chain cuprates studied here, most probably due to reduced screening in 1D.

*The two-leg ladder system: $(La,Sr,Ca)_{14}Cu_{24}O_{41+\delta}$*

The complicated system $(La,Sr,Ca)_{14}Cu_{24}O_{41+\delta}$ - which has been lovingly nicknamed the telephone-number compound by its researchers - contains both edge-sharing chains (Fig. 1a) and two-leg ladders (Fig. 1e). The interest in this system was sparked by the discovery of superconductivity at 11K under pressure for almost pure $Ca_{14}Cu_{24}O_{41+\delta}$ [15] – an observation that was remarkable for two reasons: (i) it is the first superconductivity observed in a cuprate which is not based on $CuO_2$ planes; (ii) superconductivity (or rather pairing of holes) was predicted for two-leg ladders from theoretical work[7,16] long before the telephone-number compound was first synthesised. In these systems, the number of holes can be tuned by the stoichiometry. For the '$Ca_{14}$' or '$Sr_{14}$' systems, there are 6 holes per formula unit and for $La_6(Sr,Ca)_8$, the system is formally undoped (although there is often some doping still present caused by oxygen excess).

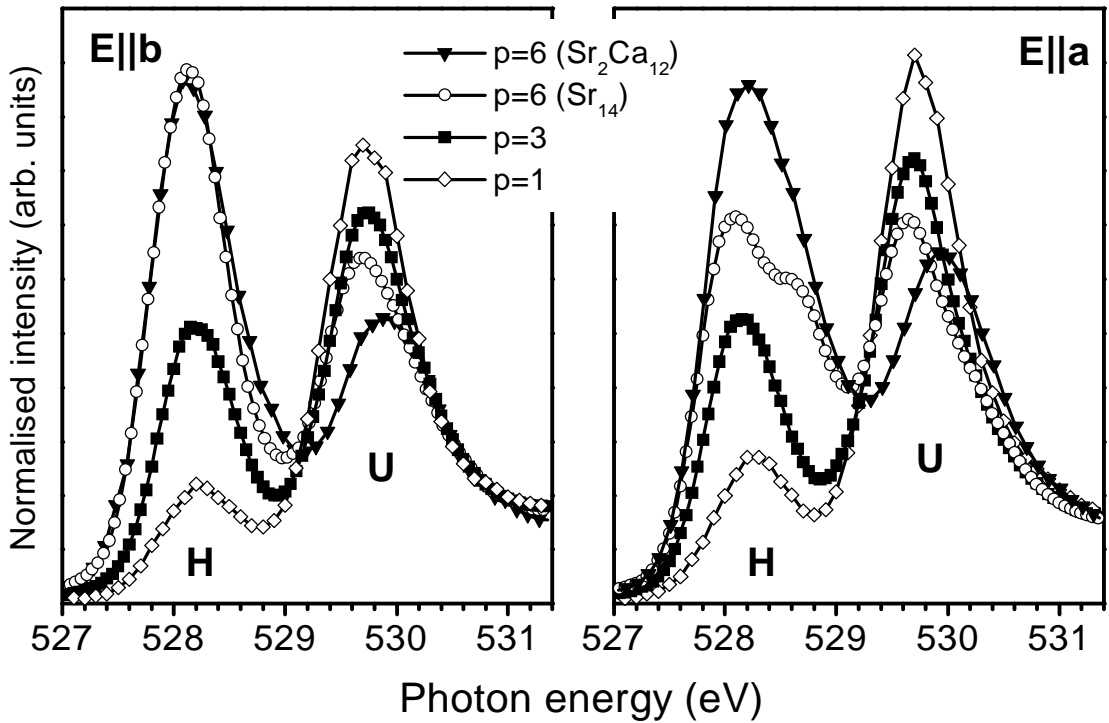

*Fig. 3.*
*Polarisation dependent XAS of the two-leg ladder system $(La,Sr,Ca)_{14}Cu_{24}O_{41+\square}$ recorded at the O-K pre-edge region [17]. Data are shown for different hole doping levels per formula unit, p [11]. p=1 refers to the compound $La_1Sr_{13}Cu_{24}O_{41}$; p=3 to $La_3Sr_{11}Cu_{24}O_{41}$. For p=6, data are shown for the Sr end-member $Sr_{14}Cu_{24}O_{41}$ (labelled $Sr_{14}$) and the almost fully Ca substituted system $Ca_2Sr_{12}Cu_{24}O_{41}$ (labelled $Ca_2Sr_{12}$).*

Fig. 3 shows selected O-K polarisation dependent XAS data[17] from the telephone-number family of compounds. The complicated crystal structure means that in the ladder plane there are 5 O2p orbitals that are σ-hybridised with $Cu3d_{x^2-y^2}$ orbitals, whereas data can be recorded for three different directions of the polarisation vector of the synchrotron radiation. Nevertheless, a careful analysis of the data allows the following picture to be built up[17]:

(i) for 1 hole and 3 holes per formula unit, a clear UHB feature is seen (E~ 529.75 eV), as well as a new pre-edge feature at lower energy (~528 eV), related to the doping induced hole states[9]. For both of these doping levels the spectra are isotropic in the ladder plane[10]. This is reminiscent of the situation for the intrinsic hole density in $Li_2CuO_2$ (Fig. 2a) and indicates that the edge-shared chains play host to the holes for these doping levels (and also for systems with p=4, independent of the Sr/Ca ratio) [17].

(ii) For the 'Sr$_{14}$' system, which is non-superconducting, there appear obvious differences between the spectra for E∥a and E∥b, in the form of a shoulder at the high energy side of the hole peak in the former. This shoulder signifies the injection of hole density into orbitals at ladder O sites - specifically the leg site is seen for E∥a. A fit to the data indicates that only one of the six doped holes resides in the ladder units in 'Sr$_{14}$' – a conclusion which confirms the 'normalisation' taken in the analysis of optical data[18].

(iii) Increasing the chemical pressure by replacing most of the Sr with Ca [giving essentially the 'Ca$_{14}$' system], leads to a change in the ratio of the lower energy component ('chains') of the hole-derived feature to the higher energy component ('ladder') for E∥b [17]. However, the level of charge transfer to the ladders upon Ca substitution for Sr is significantly lower in XAS than that proposed from optical studies of highly Ca substituted p=6 samples, where a high hole count of 2.8 on the ladders was derived[18]. The reasons for this disagreement are, at present, unclear. Although in the XAS data it is difficult, a priori, to decide whether the increase in hole density is at the rung or leg sites in the ladders, but in any case this result indicates an increase in the 2D nature of the ladder Cu-O network which may be crucial for enabling superconductivity in the highly Ca-doped systems.

The example of the telephone-number ladder compounds illustrates nicely that the site and symmetry-selective nature of XAS means that valuable information regarding the hole distribution, inaccessible via other means, can be derived in such systems, despite the complex Cu-O network geometry involved. In this context, the value of reliable 'background' information from the simple model systems – here the polarisation dependent XAS of the edge-shared CuO$_2$ chain of Li$_2$CuO$_2$ - is immediately apparent.

*The 2D planar systems Sr$_2$CuO$_2$Cl$_2$ and Ba$_2$Cu$_3$O$_4$Cl$_2$*

For completeness, we mention here the XAS results from the oxychlorides Sr$_2$CuO$_2$Cl$_2$ and Ba$_2$Cu$_3$O$_4$Cl$_2$, which are ideal model systems for undoped, 2D CuO$_2$ and Cu$_3$O$_4$ planes, as both lack apical O sites. Polarisation dependent XAS shows clearly that the vast majority of the low-lying UHB-related spectral weight is to be found in the a,b plane[19]. Furthermore, comparison of the spectra with the results from LDA band structure calculations showed the presence of hole density in O2p and Cu3d$_{3z^2-r^2}$ orbitals due to hybridisation of these levels with the empty Sr4d/Ba5d/Cu4p$_z$ and Cu4s orbitals[19]. XAS studies of the doped HTSC cuprates are dealt with in detail in Ref. [9] and will therefore not be further discussed here.

## 4. Core level photoemission of undoped 1D and 2D cuprates

In the last section, we have seen how the application of polarisation dependent XAS to single crystalline cuprates can give a wealth of information regarding correlation effects in 1-2D Cu-O networks. In this section we turn to another core level spectroscopy and deal with photoemission from core levels.

Far from being a mere tool for surface analysis, core level photoemission has matured into a refined probe of correlation effects in transition metal and rare-earth systems. The origin of the information obtainable in core level photoemission is the reaction of the valence electron system to the creation of a core hole. Through comparison of experiment with theory, such measurements offer the possibility of fixing the values of parameters entering into models for the electronic structure of strongly correlated systems.

Fig. 4 shows the experimental Cu2p$_{3/2}$ spectra for Li$_2$CuO$_2$ and Sr$_2$CuO$_3$ (symbols)[20]. The ionisation of the Cu2p level provokes a reaction of the valence electrons, with two main groups of final states for each spin-orbit component being the result. Thus both spectra show, besides the so-called main line at around 933 eV, a multiplet-split satellite feature between 940 eV and 946 eV. The satellite is due to a final state in which the core hole is poorly screened, i.e. a Cu$\underline{2p}$3d$^9$ configuration, where $\underline{2p}$ denotes a core hole in the 2p shell. The main line then corresponds to a well-screened final state, generally denoted as Cu$\underline{2p}$d$^{10}\underline{L}$, in which the core hole is screened by a charge transfer from the ligands ($\underline{L}$ - or 'ligand hole') to the core-ionised site[21]. It has been pointed out[22] that there may be different screening channels for the main line, for instance a more localised screening, in which the pushed out hole remains close to the core-ionised Cu site, and a more delocalised screening in which the hole has moved away from core hole site. Thus, through high resolution measurements of the Cu2p$_{3/2}$ main line, information on the dynamics of holes in this screening process can be obtained.

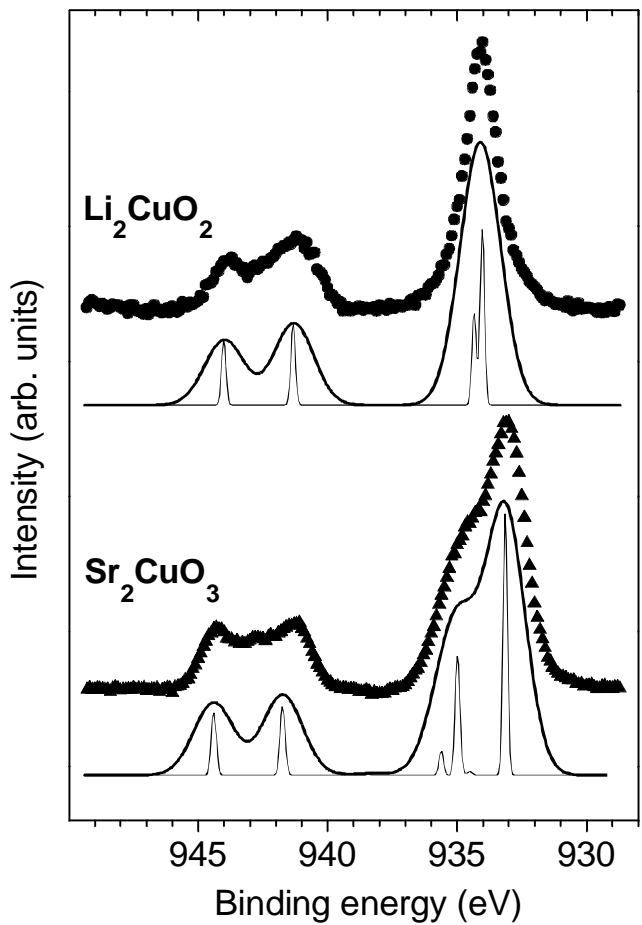

*Fig. 4.*
*High resolution Cu2p$_{3/2}$ core level photoemission spectra of Li$_2$CuO$_2$ and Sr$_2$CuO$_3$ [21]. Also shown as solid lines are calculated spectra from a three-band Hubbard model ([Li$_2$CuO$_2$: Ref. [24], Sr$_2$CuO$_3$: Ref. [23]), the results of which have either been broadened with a Gaussian of either 0.2 eV (to show the individual final states) or 1.8 eV for comparison with experiment.*

Starting with the spectrum of Li$_2$CuO$_2$, we see that the main line is composed of a single component. In this case, the nature of the Cu$\underline{2p}$3d$^{10}\underline{L}$ final state is clear. As the plaquettes in this system are edge-sharing, the interaction between plaquettes is small (see Section 2). Consequently, the ligand hole is localised predominantly on the four oxygen atoms surrounding the core-ionised copper site: in this quasi-0D system only locally screened final states are possible[20]. Also shown in Fig. 4 are simulated spectra, derived from a three-band Hubbard model using a cumulant projection technique[23]. The advantage of this method is the treatment of infinite systems, thus circumventing some of the finite size problems which limit exact diagonalisation studies.

As can be seen from Fig. 4, the spectrum of $Li_2CuO_2$ can be very well reproduced by calculations based on an *isolated* $CuO_4$ plaquette, supporting the idea discussed above that only local screening is relevant for the main line in this compound. The model parameters required[24] are $\Delta$=2.7 eV, $t_{pd}$=1.25 eV, $t_{pp}$=0.32 eV and $U_{dc}$=8.35 eV, where $t_{pp}$ is the hopping integral between O sites and $U_{dc}$ is the Coulomb interaction between the core-hole and a d electron.

Turning now to $Sr_2CuO_3$, it is clear that the main line is now composed of two features, with the main component located at lower binding energies than the locally screened final state in $Li_2CuO_2$. In the corner-shared $CuO_3$ chain of $Sr_2CuO_3$, the interaction between the plaquettes is strong and there is therefore a high probability for screening channels involving the delocalisation of the hole away from the core-ionised Cu site[20]. The calculated spectrum, in which this delocalised screening process is dominant, agrees excellently with experiment (using the parameter set $\Delta$= 2.7 eV, $t_{pd}$=1.3 eV, $t_{pp}$=0.65 eV, $U_{dc}$=8.35 eV and $U_{dd}$=8.8 eV [23]). Examination of the theoretical hole distribution in the final states shows that the dominant screening channel is one in which the ligand hole has moved to the neighbouring plaquette forming a Zhang-Rice singlet[25]. In this screening process, the total energy of the system is reduced by the stabilisation energy of a Zhang-Rice singlet and, therefore, this final state appears at the lowest binding energy. The next feature at higher binding energy corresponds to the locally screened final state in analogy to the single line seen in $Li_2CuO_2$. A similarly good agreement with the experiment for the chain systems has been reached using an approach based upon the Anderson impurity model (AIM)[26].

As regards 2D, $CuO_2$ networks, we remark that in both cluster calculations[27] and the AIM approach[26], it has proved difficult to correctly reproduce the $Cu2p_{3/2}$ spectrum, and in particular the width of the main line observed for $Sr_2CuO_2Cl_2$ [27]. It appears as though the calculations 'miss' a final state screening channel located energetically in the centre of the main line. As the oxychlorides play a special role as model substances for the HTSC, it is important to try to understand this discrepancy between theory and experiment.

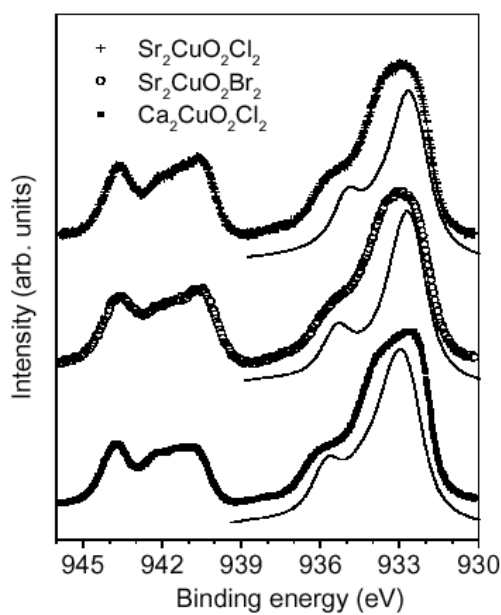

*Fig. 5*
*High resolution $Cu2p_{3/2}$ core level photoemission spectra of the $CuO_2$ plane systems $Sr_2CuO_2Cl_2$, $Sr_2CuO_2Br_2$ and $Ca_2CuO_2Cl_2$ [27,28]. Also shown as solid lines are calculated main-line spectra from an Anderson impurity model [26,28], the results of which have been broadened with a Lorentzian of width 1.1 eV and a Gaussian of 0.4 eV for comparison with experiment.*

In Fig. 5 we show $Cu2p_{3/2}$ photoemission data from the undoped, 2D $CuO_2$ planes of the oxyhalides $Sr_2CuO_2Cl_2$, $Ca_2CuO_2Cl_2$ and $Sr_2CuO_2Br_2$ [28]. The $Sr_2CuO_2Cl_2$ data are consistent with those already in the literature[27], and exhibit a very broad main line which is clearly made up of at least three components. The lowest energy part is generally agreed to be due to final states involving non-local screening, whereas the highest binding energy component of the main line is due to locally-screened final states.

The choice of substances here was guided by two ideas: (i) $Ca_2CuO_2Cl_2$ offers a test of whether the spectra from the Sr system are 'universal' from $CuO_2$-plane oxychlorides; (ii) $Sr_2CuO_2Br_2$ offers the possibility of testing the importance of the apex atom in the core-hole screening process. An apex screening channel had been suggested in the past in the context of a comparison between the $Cu2p_{3/2}$ spectra of $Nd_2CuO_4$ and $Bi_2Sr_2CaCu_2O_8$ [29].

As can be seen immediately from Fig. 5, the main line profiles of all three oxyhalides are extremely similar. For $Ca_2CuO_2Cl_2$ the intensity ratio of the satellite to main line feature is 0.38, which is significantly smaller than the value of 0.5 for both Sr systems. This is consistent with the reduced Cu-O bond length and the concomitant increase of $t_{pd}$ in the Ca compound. As a result of this we can be sure that the experimental spectra are representative of the true $CuO_2$-plane physics, as would be expected from crystals with such excellent cleavage properties giving contamination-free surfaces. The lack of any changes on swapping the apex atom from Cl to Br rules out a significant apex screening-channel, as both the Cu-apex bond lengths and electronegativities/covalency differ significantly between the two halogens.

Also shown in Fig. 5 are spectra of the main lines calculated within an AIM for each system. The comparison shows that also for $Ca_2CuO_2Cl_2$ and $Sr_2CuO_2Br_2$ the calculation significantly underestimates the width of the main line, and seems to 'miss' a screening channel in the centre. An analysis of the calculated spectra with reduced broadening[28] indicates that the high energy shoulder, which can be ascribed to final states involving host states localised near to the impurity (the core hole site), is a relatively structureless peak, whereas the low energy side of the main line (arising from final states involving delocalised host states) is more intense and rich in structure. Thus, one suggestion could be that the central intensity observed in the experiment involves delocalised host levels and could be 'missing' in the AIM due to the treatment of the hybridised O2p-Cu3d levels of the host system at the level of LDA - i.e. without taking the strong correlation effects of the other Cu3d orbitals explicitly into account. Here one is possibly confronted with the dilemma that the cluster models treat the correlation at a fundamental level, but can only deal with a limited basis set of orbitals, whereas the AIM uses the power of band structure methods to determine precisely the hopping matrix elements for the full crystal structure, but only includes correlation sufficiently at the core ionised site. Concluding this section, it is fair to say that further theoretical progress is required before we can fully understand the full complexity of the $Cu2p_{3/2}$ main line in the photoemission of model $CuO_2$-plane systems.

## 5. Angle resolved photoemission of model 2D cuprates

In this section we deal with ARPES investigations of $Sr_2CuO_2Cl_2$ and $Ba_2Cu_3O_4Cl_2$. These experiments probe the dynamics of a single hole (formed by photoemission) in the otherwise undoped $CuO_2$ or $Cu_3O_4$ plane of the system and therefore offer a touchstone for aspiring theories aiming to describe the low doping regime in the HTSC.

*ARPES of an undoped $CuO_2$ plane: $Sr_2CuO_2Cl_2$*
In Fig. 6 we show four series of ARPES spectra of $Sr_2CuO_2Cl_2$ recorded with linearly polarised synchrotron radiation of energy 22.4 eV [30]. Figs. 6(a) and 6(b) show data for $\Gamma$-$(\pi,\pi)$ and Figs. 6(c) and 6(d) for $\Gamma$-$(\pi,0)$. For both left panels [(a) and (c)], the polarisation vector of the synchrotron radiation was perpendicular to the plane formed by the emission direction of the photoelectrons and the surface normal; for both right panels [(b) and (d)], the polarisation vector lay in the emission plane. For both high symmetry directions, the emission plane is also a mirror plane of the $CuO_2$ structure: for $\Gamma$-$(\pi,\pi)$ it runs at 45° to the Cu-O bond direction and for $\Gamma$-$(\pi,0)$ along the Cu-O bonds. Thus, strictly speaking, the 'perpendicular' experimental geometry relevant for Fig. 6(a) and 6(c) selects only initial states with odd symmetry with respect to the respective mirror plane, whereas the 'parallel' geometry (Fig. 6(b) and 6(d)) selects only initial states with even reflection symmetry.

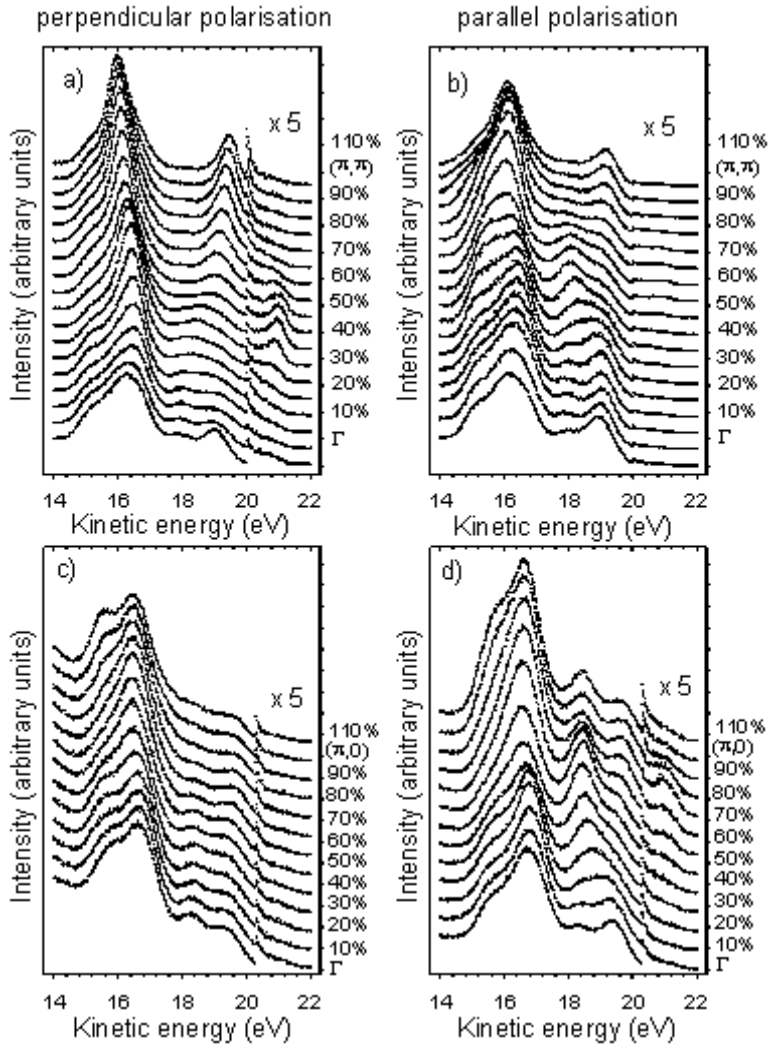

*Fig. 6*
*Series of ARPES energy distribution curves recorded from $Sr_2CuO_2Cl_2$ with linearly polarised synchrotron radiation of energy 22.4 eV [30]. The left panels, (a) and (c), show data for a geometry in which the polarisation vector is perpendicular to the plane formed by the exiting photoelectrons and the surface normal for the high symmetry directions $\Gamma$-$(\pi,\pi)$ and $\Gamma$-$(\pi,0)$, respectively. For the right panels, (b) and (d), a parallel geometry of polarisation and emission plane was chosen. In each case, the lowest lying ionisation states are displayed on an expanded y-scale and the position in k-space can be read off from the extrapolated intersection of the right hand edge of the specta with the y-axis.*

As is clear from the preceding discussion, this property of the photoemission matrix element can be used in polarisation dependent ARPES investigations to analyse the symmetry of the states involved.

The data of Fig. 6 illustrate that the first electron removal states of a $CuO_2$ plane have odd reflection symmetry with respect to $\Gamma$-$(\pi,\pi)$ and even with respect to $\Gamma$-$(\pi,0)$ - which matches the symmetry of a Zhang-Rice singlet[25]. What is also apparent from Fig. 6 is that the intensity of the first electron removal states is small compared with the main valence band states. The generally weak intensity and significant breadth of the lowest binding energy structure, coupled to the fact that one is doing photoemission on an insulator with a charge gap of some 2 eV, makes an accurate analysis of the dispersion relation of the Zhang-Rice singlet intrinsically tricky. Furthermore, as is immediately apparent from Fig. 7, the Zhang-Rice singlet intensity in ARPES is also strongly dependent on the photon energy. This is important, bearing in mind the low intensity of the first electron removal states (for example near to the high symmetry point $(\pi,0)$ in $Sr_2CuO_2Cl_2$), and severely complicates the comparison of data from earlier studies[31].

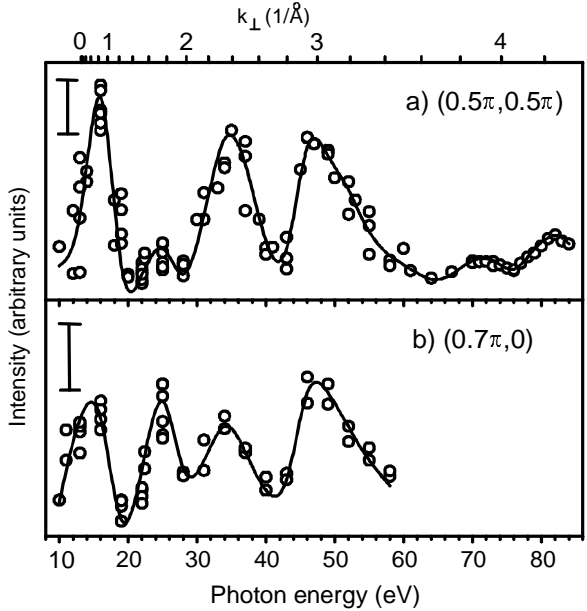

*Fig. 7
The photon energy (and corresponding $k_\perp$-) dependence of the intensity of the Zhang-Rice singlet emission in the ARPES spectra of $Sr_2CuO_2Cl_2$ for the k-points (a) ($\pi/2,0$) and (b) ($0.7\pi,0$). For each of the high symmetry directions these k-points possess the highest Zhang-Rice singlet intensity in photoemission. Note the oscillatory variation in intensity with a period of ca. $0.8 Å^{-1}$ in $k_\perp$ [30].*

Interestingly, it is also clear from Fig. 7 that the intensity of the Zhang-Rice singlet states varies with photon energy in an oscillatory manner, with a period in $k_\perp$ of ca. $0.8\ Å^{-1}$, which corresponds to a real space distance of some 7.9 Å [30]. This distance in turn corresponds to the c-axis separation of the $CuO_2$ planes, and therefore this oscillatory dependence on the photon energy can be attributed to interference effects of the photoelectron wave diffracted from the c-axis periodicity of the layered crystal structure, similar to the explanation of the strong photon energy dependence of the photoemission intensity from the molecular orbitals of $C_{60}$ [32]. The dependence of the photoemission data on both the photon polarisation and energy is a clear indication of the significant influence of matrix elements in the photoemission from these layered cuprates.

Fig. 8 shows a summary of ARPES data recorded along the two high symmetry directions from numerous cleavages with three different photon energies. The first point to note is that the observed dispersion relation does not depend on the photon energy used to record it, which reassures us that we are measuring a quantity closely related to the spectral function.

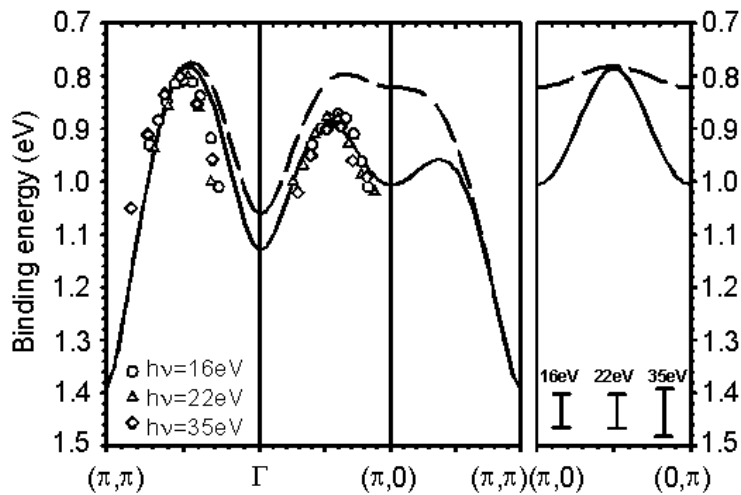

*Fig. 8
Symbols: the dispersion relation of the Zhang Rice singlet states in the undoped $CuO_2$ plane of $Sr_2CuO_2Cl_2$ from ARPES using the photon energies indicated [30]. The vertical error bars indicate the k-resolution in each case. The solid (dashed) line shows the best fit to the data within the framework of an extended (a simple) t-J model. For details see text.*

Also shown in Fig. 8 are the results of calculations carried out using the t-J and extended t-J models. It is evident that both models describe the behaviour along with the $\Gamma$ to ($\pi,\pi$) direction quite successfully, predicting parabolic dispersion centred around ($\pi/2,\pi/2$), which is the point at which the Zhang-Rice singlet has its minimum binding energy (maximum kinetic energy in Fig. 6). The remarkable thing here is the very narrow band width: of the order of 300-350 meV, describing the delocalisation of the Zhang-Rice singlets in the system. This is a result of the antiferromagnetic spin arrangement in the $CuO_2$ plane and can be visualised in the following manner. In order to let the hole delocalise in the plane, spins have to be flipped. Thus the further the hole moves, the more misaligned spins are created. Consequently, in a perfect Néel state at T=0, the holes would be completely localised. Only spin fluctuations, determined by the exchange integral J, enable the frustrated spins to flip back to match the antiferromagnetic background. This is the reason why the dynamics of the holes in such insulator systems are not determined by the hopping integral t but by the exchange integral J, which is of the order of 120 meV.

It is along the $\Gamma$-$(\pi,0)$ direction that significant differences between experiment and theory occur. In the simple t-J model (see Fig. 8, dashed line: J=t=0.28 eV [30]), the minimal binding energies along $\Gamma$ -$(\pi,\pi)$ and $\Gamma$-$(\pi,0)$ are almost equivalent. In the experimental data there is an energy difference of some 70-100 meV between the lowest lying ionisation states along the two high symmetry directions[30]. This characteristic can be well described by the results of an extended t-J model, as is also shown in Fig. 8 as a solid line. In this case, the hopping to the second ($t_2$) and third ($t_3$) neighbours is also included, in addition to the dominating nearest neighbour hopping $t_1$. For the good fit to the data presented in Fig. 8, the following parameters were taken: J=$t_1$=0.22 eV, $t_2/t_1$=-0.1 and $t_3/t_1$=0.2 [30].

A further area in which the comparison of experiment and theory can be enlightening is that of the k-dependence of the spectral weight of the ARPES features. However, as a result of space considerations we refer the reader for details on this subject to Ref. [30].

*ARPES of an undoped $Cu_3O_4$ plane: $Ba_2Cu_3O_4Cl_2$*

We now turn to analogous measurements of the dispersion of a single hole in the $Cu_3O_4$ plane of the model cuprate $Ba_2Cu_3O_4Cl_2$. As mentioned in Section 2 (see also Fig. 1f), the two subsystems into which this plane can be split have radically different $T_N$'s, thus offering the unique possibility of investigating the dispersion of a single hole in both an antiferromagnetic and paramagnetic spin background simultaneously[33]. Performing ARPES measurements at or slightly above room temperature, the hole can either be injected into the essentially antiferromagnetically ordered $Cu_A$ sublattice or the paramagnetic subsystem of the $Cu_B$ spins. These, then, represent the limiting cases as regards the doping dependence of the spin background in the HTSC.

Fig. 9 shows examples of ARPES spectra recorded along the $\Gamma$-$(\pi,0)$ direction of the $Cu_3O_4$ system[34], with the polarisation geometry such that we are sensitive to initial states of odd symmetry with respect to the mirror plane which runs at 45° to the Cu-O bonds. Here the first electron removal states have lowest binding energy and maximal intensity at $(\pi,0)$ and $(3\pi,0)$, which are the equivalent points to $(\pi/2,\pi/2)$ in $Sr_2CuO_2Cl_2$. However, after passing $(\pi,0)$, their intensity does not fall away sharply as after $(\pi/2,\pi/2)$ in $Sr_2CuO_2Cl_2$, but remains strong right up to $(3\pi,0)$ [33]. This extra intensity is that of the photohole in the paramagnetic $Cu_B$ sublattice, whereas the spectral weight at lowest binding energies around $(\pi,0)$ is that of the photohole in the antiferromagnetically ordered $Cu_A$ sub-lattice.

From the data shown in Fig. 9, as well as a wealth of other data recorded with different photon energies and along other directions in k-space[33], we can derive the experimental dispersion relations for the Zhang-Rice singlet dispersion on the $Cu_A$ and $Cu_B$ subsystems and compare them with the results of model calculations, as shown in Fig. 10.

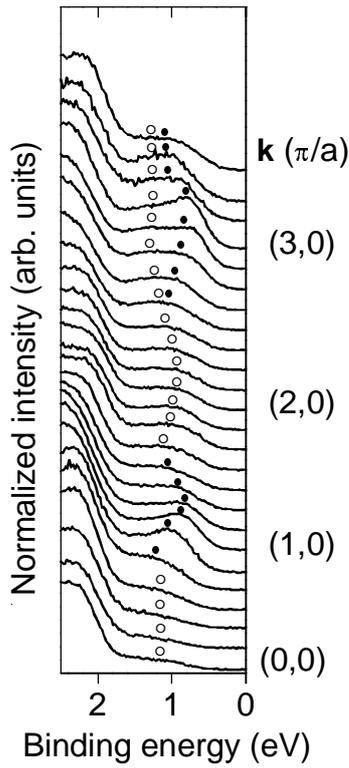

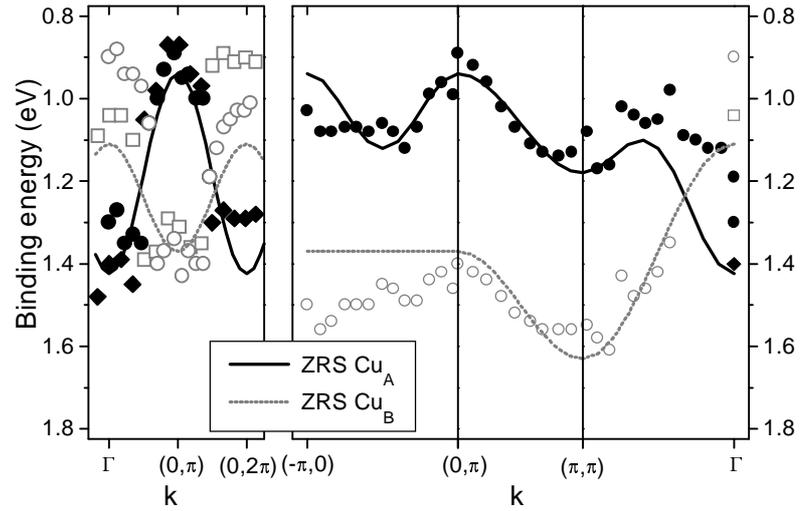

Fig. 9 (left).
A series of ARPES energy distribution curves recorded from the undoped $Cu_3O_4$ plane of $Ba_2Cu_3O_4Cl_2$ using linearly polarised synchrotron radiation of energy 35 eV in a perpendicular polarisation geometry [33] for k along $\Gamma$-$(\pi,0)$ [34]. The symbols indicate the dispersion of the Zhang-Rice singlet on the $Cu_A$-O (●) and $Cu_B$-O (○) subsystems.

Fig. 10 (right).
Comparison of the dispersion relation of the Zhang-Rice singlet on the $Cu_A$-O (●◆) and $Cu_B$-O (○□) subsystems in the undoped $Cu_3O_4$ plane of $Ba_2Cu_3O_4Cl_2$ measured in ARPES (●○:hν=35 eV; ◆□: hν =20 eV), with calculations describing the J-mediated Zhang-Rice singlet propagation of a hole in an antiferromagnetic spin background on the $Cu_A$-O subsystem (black solid line) and the t-mediated propagation of a hole in the paramagnetic $Cu_B$-O subsystem (grey dotted line) [33].

The $Cu_A$ data (experiment: black symbols; theory: black solid lines) result in a bandwidth determined by J, in analogy with the observed dispersion in the $CuO_2$ planes of $Sr_2CuO_2Cl_2$. A J-value of 0.22 ± 0.03 eV is arrived at[33], which is of the correct magnitude. The hole in the paramagnetic spin background ($Cu_B$ subsystem), however, follows a tight-binding-like dispersion relation (experiment: grey symbols; theory: grey dotted lines) with a hopping integral $t_B$ = -0.13 ± 0.05 eV. The value of $t_B$ agrees roughly with the estimated hopping integral between the $Cu_B$ sites[33].

In summary, these data illustrate nicely the strong influence of the spin environment on the dynamics of holes in the cuprates. In particular, as far as the spin background is concerned, the ARPES data from the $Cu_3O_4$ planes of $Ba_2Cu_3O_4Cl_2$ represent simultaneously both the low and very high doping limits in 2D cuprate materials.

To close this section we point out that it is interesting to compare the dynamics of the single-particle excitations measured in ARPES experiments such as those presented here with the q-dependence of electron-hole excitations as measured using electron energy-loss spectroscopy in transmission[35]. For example, in both the 1D and 2D antiferromagnets $Sr_2CuO_3$ and $Sr_2CuO_2Cl_2$, dispersive band widths of the two-particle excitations across the charge transfer gap of the order of 1-1.5 eV are observed. This illustrates that a spin-less entity such as an electron-hole pair can have radically different dynamics than a single hole injected into the same Cu-O network.

## 6. X-ray absorption of hole doped edge-shared cuprate chains

The preceding sections have dealt exclusively with undoped, model cuprates. In the closing two results-based sections of the paper we deal with some new data from 1D and 2D hole-doped cuprates.

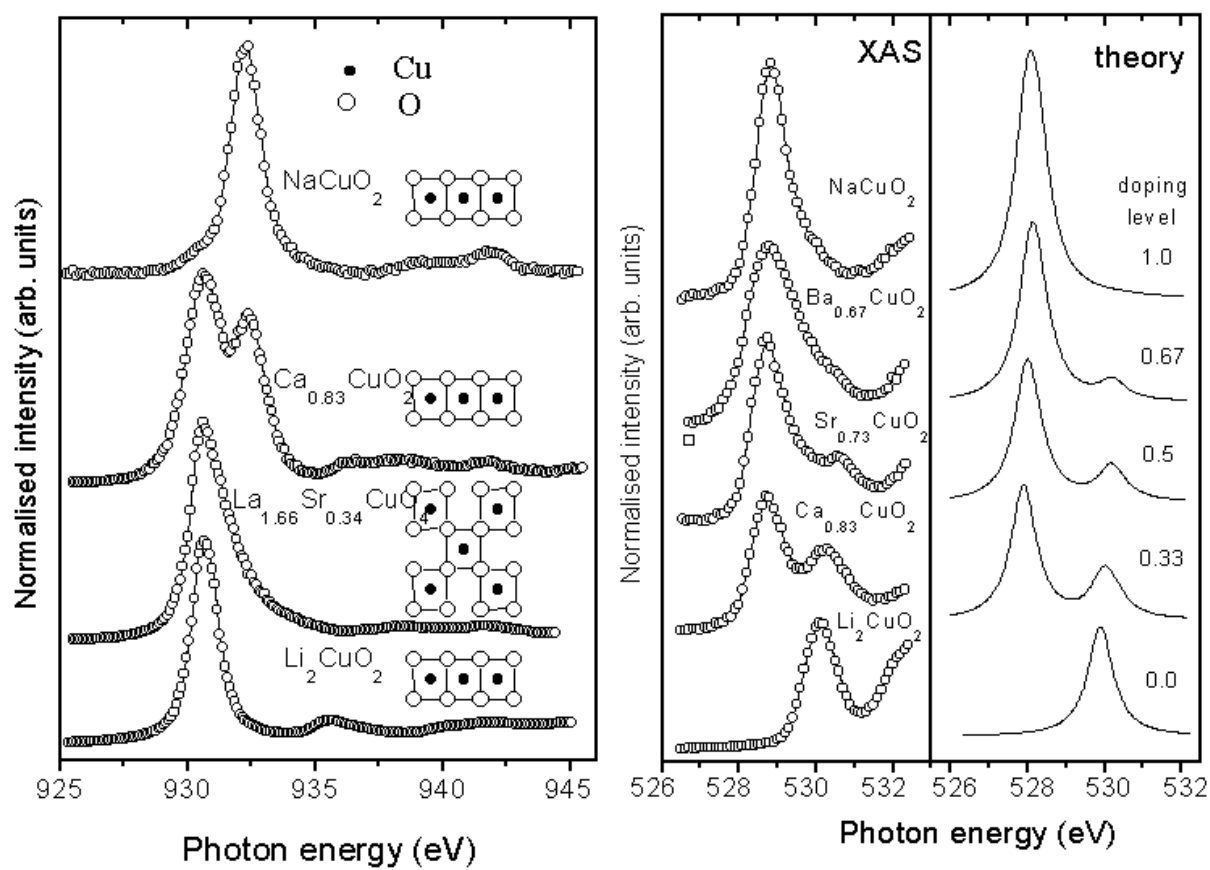

*Fig. 11*
*Left panel: x-ray absorption at the Cu-$L_3$ edge of the systems as indicated. The sketches show the Cu-O network geometry in each case. Note the significantly enhanced $\underline{2p}3d^{10}\underline{L}$ final state feature at 933 eV in the case of $Ca_{0.83}CuO_2$ ($Cu^{2.34+}$), in comparison to the spectrum of the equally doped $La_{1.86}Sr_{0.34}CuO_4$ system [36]. Right panel: (lines and symbols) x-ray absorption at the O-K pre-edge region of the edge-shared chain systems as indicated. The results of an extended 5-band Hubbard model for an edge-shared chain with a hole doping level per Cu site as given are shown as solid lines for comparison on the right [36].*

Fig. 11 shows x-ray absorption data from a series of edge-shared chain compounds, in which the doping level has been varied from zero to one hole per Cu site[36]. The left part of the figure contains the Cu-$L_3$ spectra and the right part the O-K edges, together with the results of a cluster calculation of the XAS data. We start with the Cu-$L_3$ spectra. That of $Li_2CuO_2$ shows the 'white line' ($\underline{2p}3d^{10}$ final state) associated with the single intrinsic 3d hole present in the ground state of this formally divalent system. The uppermost spectrum, from the Cu(III) analogue $NaCuO_2$, is also comprised of a single line, which now represents essentially a $\underline{2p}3d^{10}\underline{L}$ final state. In between these two end-members we show the Cu-$L_3$ spectra of two different Cu-O networks with the same formal doping level of $Cu^{2.34+}$. The spectrum from $La_{1.86}Sr_{0.34}CuO_4$ resembles the vast majority of Cu-$L_3$ spectra of doped HTSC[9], with a slight asymmetry due to the ligand-hole final state visible on the high energy side of the $\underline{2p}3d^{10}$ white line. The situation in $Ca_{0.83}CuO_2$, however, is radically different. The Cu-$L_3$ spectrum in this case resembles a weighted sum of the two end-member spectra, i.e. the relative intensities of the $\underline{2p}3d^{10}$ and $\underline{2p}3d^{10}\underline{L}$ peaks is 0.66:0.34.

The O-K edges of the hole-doped edge-shared chain systems also show a new behaviour compared to the doped 2D cuprates. As can be seen from the right panel of Fig. 11, cluster calculations based upon an extended five-band pd model describe the O-K edges very well, enabling extraction of parameters which describe the electronic structure of these systems[36]. In addition, the upper Hubbard band feature, located around 530 eV, remains visible in the spectra even up to the extremely high doping level of 0.67 holes per Cu ($Ba_{0.67}CuO_2$)[36]. In the 2D Cu-O planes of the HTSC[9], the UHB feature has already disappeared from the O-K XAS spectra for a doping level lower than that corresponding to $Ca_{0.83}CuO_2$. The survival of the UHB into the highly doped regime here is a direct consequence of the 90° Cu-O-Cu interaction pathway in the edge-shared chain systems, which strongly suppresses the inter-plaquette hybridisation. In Section 4, we saw how this leads to a behaviour analogous to that of an isolated $CuO_4$ plaquette in the screening response of the system to the creation of a core-hole in photoemission. The same holds true for the q-dependent collective excitations of $Li_2CuO_2$ measured using EELS in transmission[24,35]. What is seen in Fig. 11 is that the reduction of the inter-plaquette hopping also has consequences for the electron addition experiment (XAS): leading to a suppression of the dynamical spectral weight transfer from the high energy scale (upper Hubbard band) to the low energy scale (the hole-doping peak)[36]. As the dynamical spectral weight transfer itself is one of the hallmarks of a correlated charge transfer insulator[9], these results illustrate nicely the interplay between such correlation effects and delocalisation (described by the band width) - in this case as a function of the dimensionality and interconnection of the Cu-O network concerned.

## 7. Angle resolved photoemission of high temperature superconductors

For the last section we briefly touch on the subject of the electronic structure of the $Bi_2Sr_2CaCu_2O_{8+\delta}$ (Bi-2212) family of high temperature superconductors as investigated using ARPES. Bi-2212 can certainly lay claims to being the most intensively ARPES-studied solid in the history of this high energy spectroscopy, and much has been learned regarding the electronic structure of the HTSC from its study, as is nicely illustrated in the articles by Shen *et al.* and Campuzano *et al.* in this volume.

The question as to how the antiferromagnetic, insulating $CuO_2$ plane (as is realised in $Sr_2CuO_2Cl_2$) evolves with hole doping to give the metallic Fermi surface seen in the Bi-2212 remains one of the key questions in HTSC physics. As shown above, ARPES investigations of undoped, model cuprates can shed light mainly on the extreme low-doping limit in the canonical (T,p) phase diagramme of the cuprates - illustrating, for example, the importance of the spin background in determining the dynamics of holes in these systems. One ARPES study has attempted to connect the undoped and underdoped regimes: linking the form of the ZHANG-RICE SINGLET dispersion along the antiferromagnetic zone boundary [($\pi$,0) to (0,$\pi$)] in $Ca_2CuO_2Cl_2$ to the d-wave-like normal state pseudogap observed above $T_c$ in underdoped HTSC[37].

Naturally, in trying to bridge the 'doping-gap' in our understanding of the HTSC it is equally vital to possess robust and detailed experimental information regarding the normal state electronic structure and Fermi surface of the HTSC themselves.

In this final section, we give a flavour of the depth of information available from ARPES experiments of the HTSC carried out using 'new generation' electron energy analyser instrumentation. These machines enable photoemission to be carried out in an angle-scanned mode[38], but including the measurement of thousands of energy distribution curves per Brillouin zone quadrant, each recorded with both high angular (±0.3°) and energy (30 meV) resolution. In this way, we combine the benefits of both the angle-scanning and EDC-based methods[39]. Particularly important here is the analysis (both technically by the machine, and also by the scientist looking at the data) of the momentum distribution of the photoelectrons on an equal footing with their energy distribution. A direct experimental picture of the Fermi surface topology is then given by the shape formed by joining the maxima in a 2D network of normalised momentum distribution curves, or MDC's, each recorded for $E = E_F$ [39].

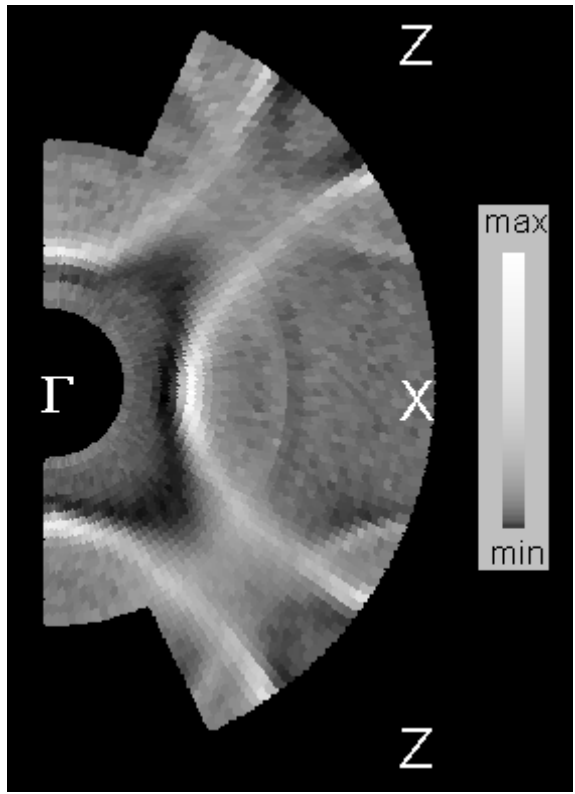

*Fig. 12.*
*Momentum distribution map with $E = E_F$ for Pb-doped Bi-2212 measured at room temperature using monochromatised He I radiation with a low degree of polarisation [40]. The bright features show the topology of the normal state Fermi surface of this system, which is clearly comprised of a main Fermi surface (hole-like, centred at the X,Y points) and shadow Fermi surface segments. For details, see text.*

Fig. 12 shows such a momentum distribution map for $E=E_F$, which can alternatively be called a Fermi surface map. The data were recorded at room temperature using monochromatised He I radiation from Pb-doped Bi-2212 [40]. The map is built up from more than $10^3$ high resolution spectra per Brillouin zone quadrant, making a subsequent massage of the data (interpolation, differentiation etc.) superfluous. An additional important advantage here is the use of radiation with a *low* degree of polarisation. The data shown in Fig. 6 from $Sr_2CuO_2Cl_2$ illustrated the impact of the photon polarisation on the measured Zhang-Rice singlet photoelectron intensity from the $CuO_2$ plane. The same holds true for Bi-2212 [41], meaning that maps recorded with highly polarised synchrotron radiation necessarily present only a part of the picture[42]. As the aim of such measurements is to attain as unbiased a view of a wide k-space region as possible, the advantage of low-polarised radiation is clear.

Fig. 12 shows the presence of two features: the main Fermi surface, which is hole-like, and has the form of barrels centred around the X,Y points and the so-called shadow Fermi surface, a part of which one can see clearly in the lower right corner of the figure. The main Fermi surface topology agrees with the vast majority of earlier ARPES data[43], as well as some recent high resolution mapping investigations[39,44], but is in clear disagreement with a few recent mapping studies carried out with highly polarised synchrotron radiation[45]. The reason for the controversy lies, we believe, in a combination of matrix element effects and the complications resulting from a host of extrinsic features in the Fermi surface maps of pure Bi-2212 [40], the latter arising from diffraction of the photoelectrons from the incommensurate b-axis modulation. This modulation is not present in Pb-doped Bi-2212, thus making this material in fact the ideal substance for ARPES investigations in the future[39,40]. We mention in passing that the FS topology we see in ARPES is fully consistent with that derived from analysis of the charge carrier plasmon in Bi-2212 as measured by EELS in transmission[35].

The very first angle-scanned photoemission investigations of Bi-2212 provided the first evidence for shadow Fermi surface features[38], which were explained in terms of the continued existence of short-range antiferromagnetic correlations[46], even deep into the hole-doped part of the HTSC phase diagramme. The data shown in Fig. 12, as well as new data from the Pb-doped Bi-2212 system (not shown), illustrate the important point that the shadow Fermi surface does not appear to be merely a copy of the main Fermi surface, with the same intensity distribution around the 'barrel' merely shifted as a whole by the antiferromagnetic wavevector $(\pi,\pi)$. The discontinuity of the shadow Fermi surface feature (it does not cut the main Fermi surface and cross the $\Gamma$-M-Z line) means that we can rule out that it can have a simple diffraction-based origin in Pb-doped Bi-2212, for example as the result of a c(2x2) reconstruction of the surface.

These results, apart from confirming, with unprecendented clarity, the hole-like nature of the main Fermi surface in the Bi-2212 systems, also indicate that the complex and fascinating interplay between charged excitations and the spin background, which itself is the result of the effects of strong electronic correlations, covers the whole doping range in the cuprate systems. Future angle-scanned photoemission measurements of the Fermi surface topology and shadow features as a function of the doping level and temperature will surely deliver important information regarding the transition in the HTSC systems with doping from the paradigm of a lightly doped charge insulator to the relatively 'normal' metallic behaviour in the overdoped region of the phase diagramme.

## 8. Conclusions

In this contribution we have briefly treated the experimental investigation of the electronic structure of cuprates as investigated using high energy spectroscopies. The latter are extremely powerful probes of correlation effects in the cuprates, providing direct experimental access to:

(i) the distribution of both intrinsic and 'doped' holes between the different orbitals of the structure from x-ray absorption spectroscopy;
(ii) the dynamics of the valence band electron system as measured by the screening response to the creation of a core hole in photoemission and
(iii) the dispersion relation of the lowest lying ionisation states in undoped cuprates and Fermi surface topology in the HTSC from angle resolved photoemission.

In combination with widely differing theoretical calculations (e.g. 1 band or 3 band Hubbard models; Anderson impurity models or band structure calculations), such experimental data enable a detailed picture to be built up of the electronic states in these systems. In particular the role played by the dimensionality of the Cu-O network concerned and the nature of the spin-background supported by that network are shown to be crucial for the understanding of the nature and dynamics of charge carriers in cuprate systems.

The ultimate goal of a widely accepted microscopic mechanism for high temperature still eludes us, but the journey thus far has brought many fascinating glimpses of the richness and complexity the cuprates have to offer, much of which is thanks to the effects of strong electronic correlation.

## 9. Acknowledgements

Financial support from the BMBF (13N 6599/9 and 05 605 BDA / 05 SB8 BDA6), the DFG (SFB 463; Graduiertenkolleg "Struktur und Korrelationseffekte in Festkörpern" der TU Dresden; FI439/7-1), and the HCM program of the EU is gratefully acknowledged. We thank H. Berger, S. Uchida, N. Motoyama, H. Eisaki, J. Karpinski, L. L. Miller, G. Yang and S. Abell for providing high quality single crystals. Experimental collaboration with G. Reichardt, C. Janowitz, R. Müller, R. L. Johnson, G. Kaindl, M. Domke, N. Nücker, S. Schuppler, and K. Maiti is acknowledged. Finally we thank S.-L. Drechsler, J. Málek, R. Hayn, W. Stephan, K. Penc, C. Waidacher, K. W. Becker, K. Karlsson, O. Gunnarsson, O. Jepsen and V.Y. Yushankhai for invaluable theoretical support and S. Atzkern for a critical reading of the manuscript.